\documentstyle[twoside,epsfig,psfig]{article} 
 
\input ibvs2.sty 
 
\begin{document} 
\IBVShead{6139}{5 Mar 2015} 
 
\IBVStitle{Photometric evolution and peculiar dust formation}
\vskip -0.3cm
\IBVStitle{in the gamma-ray Nova Sco 2012 (V1324 Sco)}
\IBVStitle{} 
  
\IBVSauth{U. Munari$^1$, F. M. Walter$^2$, A. Henden$^3$, S. Dallaporta$^4$,
T. Finzell$^5$, L. Chomiuk$^5$} 

\IBVSinst{INAF Osservatorio Astronomico di Padova, Sede di Asiago, I-36032 Asiago (VI), Italy} 
\IBVSinst{Dept. of Physics and Astronomy, Stony Brook University, Stony Brook, NY 11794-3800, USA} 
\IBVSinst{AAVSO, 49 Bay State Rd. Cambridge, MA 02138, USA}
\IBVSinst{ANS Collaboration, c/o Astronomical Observatory, 36012 Asiago (VI), Italy} 
\IBVSinst{Dept. of Physics and Astronomy, Michigan State Univ., 567 Wilson Road, East Lansing, MI~48824-2320,~USA}

\IBVStyp{Nova} 
\IBVSkey{photometry} 
\IBVSabs{Optical (BVRI) and infrared (JHK) photometry of the gamma-ray nova
Nova Sco 2012 (V1324 Sco) is presented and the lightcurve reconstructed and
discussed. An interstellar reddening E(B-V)=1.23 is derived. Dust begun to
form at an early date in the nova, only one magnitude down and 20 days past 
maximum optical brightness and caused an extinction of at least 6 magnitudes
in $V$, that cleared some months later. This unusual early dust formation
compromises the application of the magnitude at maximum versus rate of
decline (MMRD) relations in estimating the distance to the nova.}

\begintext 
Nova Sco 2012 was discovered on 2012 May 22.80 UT by Wagner et al. (2012) as
the optical transient MOA 2012 BLG-320 during the Microlensing Observations
in Astrophysics (MOA) survey (Abe et al.  1997).  It appeared at equatorial
coordinates RA = 17:50:53.90 and DEC = $-$32:37:20.46 (J2000), corresponding
to Galactic coordinates $l$=357.4255 and $b$=$-$02.8723 deg, and has no
obvious counterpart on DSS plates.  Pre-outburst MOA photometry reported by
Wagner et al.  (2012) shows the progenitor varying around 19.0-19.5 mag in
$I$ band.  Between May 14 and 16 UT, the source began a slow monotonic
increase in brightness, modulated with an amplitude of about 0.1 mag and a
period of about 1.6 hr, with the real outburst starting between June 1.77
and 2.55 UT.  High-resolution spectroscopy obtained on June 4.08 UT by
Wagner et al.  (2012) with the ESO-VLT+UVES telescope confirmed the
transient to be an FeII-class nova, with a FWHM$\sim$800 km/s for the
H$\alpha$ and H$\beta$ lines.

The Fermi/LAT satellite detected $\gamma$-ray emission at $>$100 MeV from
Nova Sco 2012 from June 16 to June 30, with the highest flux during June
18-24 (Cheung et al.  2012, Metzger et al.  2015).  Only a very few other
novae have been so far detected in $\gamma$-rays (Ackermann et al.  2014).

No description of the optical and infrared photometric evolution of Nova Sco
2012 has been provided to date.  In this paper we present $B$$V$$R_{\rm
C}$$I_{\rm C}$$J$$H$$K$ photometry of Nova Sco 2012 and discuss its
lightcurve covering the evolution from pre-maximum to day +34 past optical
maximum, when the nova had declined by more than 5 magnitudes below $V$
maximum, plus some later $B$,$V$ data.  Our photometry is given in Tables~1
and 2.  It was collected with ANS
Collaboration telescope 036 (Munari et al.  2012), during the APASS all-sky
survey (Henden et al.  2012), with AAVSOnet OC61 telescope (Mt.  John
University Observatory, NZ), and with the CTIO SMARTS 1.3 m telescope
(Walter et al.  2012).  The light- and color-curves of Nova Sco 2012 are
presented in Figure~1, with pre-maximum data from Wagner et al.  (2012) and
observations retrieved from VSNET and AAVSO international databases
included.  The AAVSO data are presented as the mean nightly value for a
single observer when multiple entries are present.  AAVSO and VSNET data
have had obviously deviating points removed, and offsets have been applied
to bring their zero points to agree with ANS, APASS and SMARTS properly
calibrated data (0.15, $-$0.05 and $-$0.17 mag have been added to VSNET $B$,
$V$ and $R_{\rm C}$ data, respectively, and 0.05 mag to AAVSO $B$ data).  At
8 arcsec distance from the nova lies a field star, measured by APASS at
$B$=16.23, $V$=15.32, $g'$=15.69, $r'$=14.95, $i'$=13.05.

\IBVSfig{17.6cm}{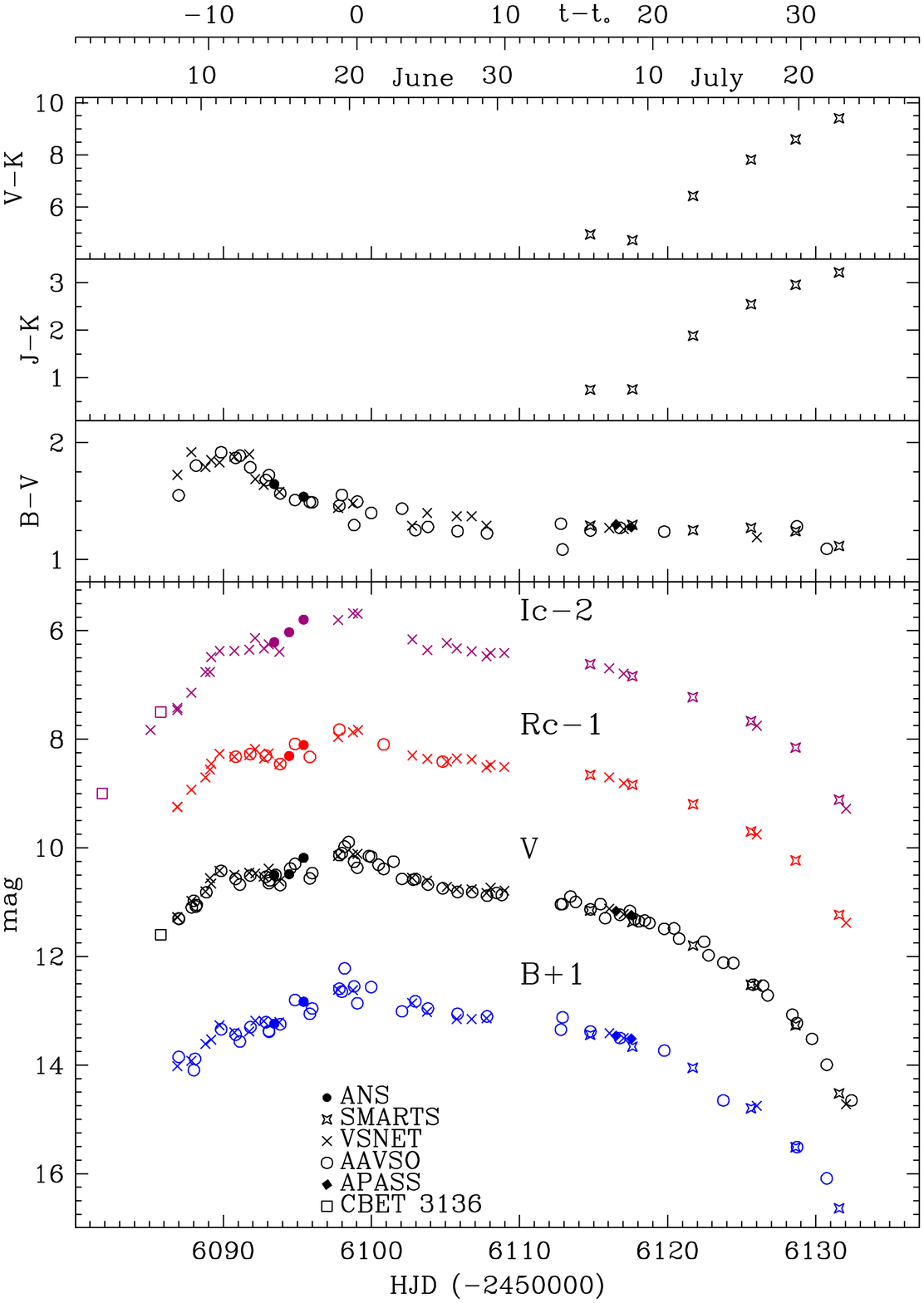}{$B$$V$$R_{\rm C}$$I_{\rm C}$ lightcurves and
$B-V$, $J-K$ and $V-K$ evolution of Nova Sco 2012.}

The rise of Nova Sco 2012 toward maximum in the $I_{\rm C}$ band covered the
last 2.65 mag before the pre-max halt at a constant speed of 0.33 mag/day. 
The pre-max halt in Figure~1 is obvious in $V$, $R_{\rm C}$ and $I_{\rm C}$
and lasted $\sim$3 days (from day $-$9 to $-$6), after which the final
$\sim$0.5 mag rise to maximum was completed in $\sim$6 days with
superimposed large variability.  The presence of the pre-max halt is much
less obvious in $B$ band.

\vskip 0.6 cm
\centerline{Table 1. Optical photometry of Nova Sco 2012.}
\vskip 3mm
\centerline{\psfig{file=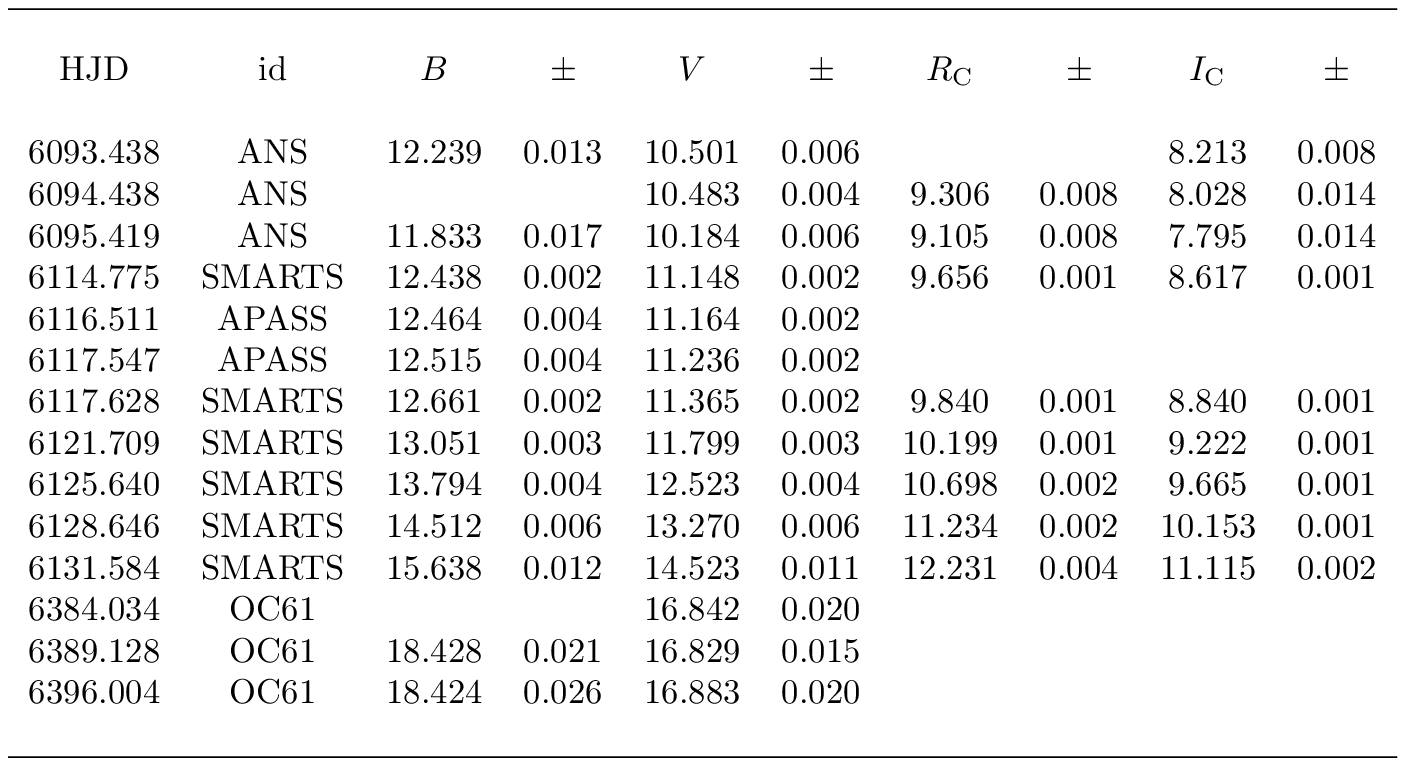,width=12.1cm}}

\vskip 0.6 cm
\centerline{Table 2. SMARTS infrared photometry of Nova Sco 2012.}
\vskip 3mm
\centerline{\psfig{file=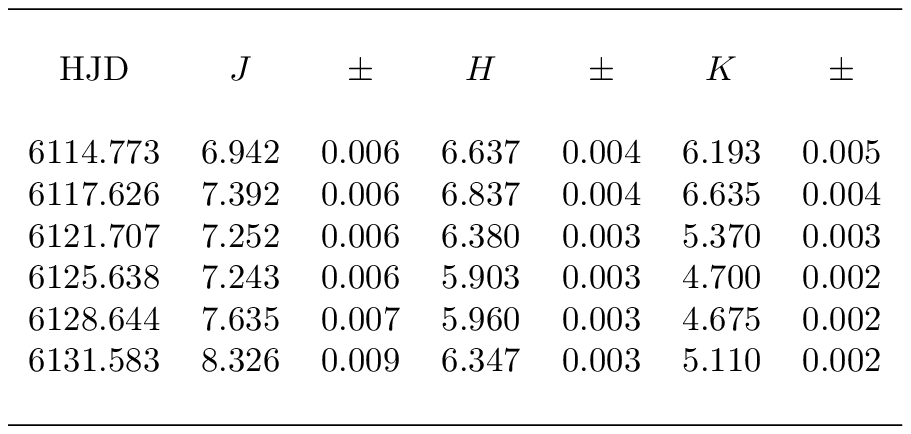,width=8.2cm}}

The initial decline from maximum of Nova Sco 2012 was smooth and slow, with
decline times of $t_{1}^{B}$=19, $t_{1}^{V}$=17 days, and a decline slope
suggesting $t_{2}^{V}$$\approx$40 days ($t_n$ is the time required to
decline from maximum by $n$ magnitudes in the given band).  Initial
$J$$H$$K$ photometry (cf Figure~1, data points for days +16 and +19) is
consistent with normal photospheric emission under heavy interstellar
reddening.  Around day +20, dust begun to form in the ejecta, which caused a
rapid and large increase in $J-K$ and $V-K$ color indexes and a parallel
drop in optical brightness.  It is worth noticing that, as observed in
many novae (e.g.  Nova Aql 1993 by Munari et al.  1994; Nova Sct 2009 by Raj
et al.  2012; Nova Cep 2013 by Munari et al.  2014), the dust condensation
did not cause a reddening of optical colors of Nova Sco 2012 (cf Figure~1). 
The fact that the wavelength-dependent absorption efficiency of the dust
turned from neutral to selective at $\lambda$$\geq$6000~\AA\ suggests a
prevalent carbon composition with a diameter of dust grains of the order of
0.1 $\mu$m (Draine and Lee 1984; Kolotilov et al.  1996). In many novae
forming dust, the dust condensation starts quite suddenly causing a marked
knee in the lightcurve (cf the recent case of Nova Cep 2013, Munari et al. 
2014).  Such a knee is absent in the lightcurve of Nova Sco 2012 and
substituted by a gradual {\em acceleration} of the extinction, suggesting a
more gradual condensation of the dust grains.  Nova Sco 2012 differs from
many other dust condensing novae in two other respects: (1) dust formation
usually occurs at the time of the transition from optically thick to
optically thin ejecta, about 3-4 mag below maximum brightness, which led
Shore and Gehrz (2004) to suggest that it is the photo-ionization from the
central star that triggers dust grain condensation in the ejecta.  In Nova
Sco 2012, dust begun to condense at a much earlier phase, only 1 mag below
maximum; (2) the slower a nova, the later dust begins to condense, as
illustrated by the compilation of data for M31 and Galactic novae of Shafter
et al.  (2011).  Their relation, applied to the predicted
$t_{2}^{V}$$\approx$40 days for Nova Sco 2012, indicates that dust should
have begun to condense $\sim$60 days past maximum ($\pm$10 days given the
dispersion of the plotted data) and $\sim$60 days is also estimated from the
theoretical modeling presented by Williams et al.  (2013).  Such a $\sim$60
day delay is much longer than the observed 20 days for Nova Sco 2012.  The
dust extinction caused the $V$ magnitude of Nova Sco 2012 to drop below
19-20 mag (SMARTS and AAVSO data).  When we re-observed the nova in April
2013 (day +285, see Table~1) the nova had rebrightened to $V$=16.8 and
recovered the normal brightness decline, meaning the dust extinction had
already cleared.

According to van den Bergh and Younger (1987), typical novae display
($B$$-$$V$)$_\circ$=+0.23 at $B$,$V$ maximum brightness and
($B$$-$$V$)$_\circ$=$-$0.02 at $t_{2}$.  The corresponding values for Nova
Sco 2012 were $B$$-$$V$=+1.44 and $B$$-$$V$=+1.23 (from spline interpolation
of the data in Figure~1), that provide $E_{B-V}$=1.21 and 1.25,
respectively.  The average $E_{B-V}$=1.23 is in excellent agreement with the
$E_{B-V}$=1.23 derived by Finzell et al.  (2015) from the intensity of
interstellar NaI and KI in high resolution spectra of Nova Sco 2012.  The
corresponding extinction is $A_V$=4.18 mag from Fiorucci and Munari (2003)
relations appropriate for the intrinsic colors of the nova and standard
$R_V$=3.1 extinction law.

Nova Sco 2012 reached a maximum brightness of $B$=11.60, $V$=10.14, $R_{\rm
C}$ = 8.85 and $I_{\rm C}$ = 7.68 around JD=2456099.0 ($\pm$0.5), 2012 June
20.5 UT, and the observed decline times were $t_{2}^{B}$=26, $t_{2}^{V}$=25,
$t_{3}^{B}$=30.5, $t_{3}^{V}$=29.5, with $t_3$ times too short with respect
to $t_2$ because of the undergoing extinction by dust condensing in the
ejecta.  These decline times are a popular means to estimate the distance to
a nova.  The most recent calibration of the absolute magnitude at maximum
versus rate of decline (MMRD) relation has been presented by Downes and
Duerbeck (2000).  The distance to Nova Sco 2012, with the $A_V$=4.18 mag
extinction, is 5.5 kpc for $t_{2}^{V}$=25 and the linear MMRD and 5.3 kpc
for the S-shaped MMRD.  For $t_{3}^{V}$=29.5, the linear MMRD yields 6.9
kpc.  The large discrepancy of the distances estimated from $t_{2}^{V}$ and
$t_{3}^{V}$ is a sign of the disturbance from the condensing dust. 
Considering $t_{2}^{V}$$\approx$40 days, estimated above as a feasible value
in absence of condensing dust, the corresponding distance to Nova Sco 2012
would be 4.3 kpc from linear MMRD and 3.7 kpc from S-shaped MMRD.

\references 

  Abe, F., et al. 1997, in "Variables Stars and the Astrophysical Returns of the Microlensing Surveys", R. Ferlet, J.-P. Maillard and B. Raban eds., Editions Frontieres, p.75 

  Ackermann, M., et al.,  2014, Science 345, 554

  Cheung, C. C., et al., 2012, ATel 4310

  Downes, R.~A., Duerbeck, H.~W., 2000, AJ, 120, 2007

  Draine B. T., Lee H. M., 1984, ApJ, 285, 89

  Finzell, T., et al., 2015, in preparation

  Fiorucci, M., Munari, U.,  2003, A\&A 401, 781

  Henden, A.~A., et al., JAAVSO 40, 430

  Kolotilov E. A., et al., 1996, ARep., 40, 81

  Metzger, B.~D., et al., 2015, MNRAS in press (arXiv 1501.05308) 

  Munari, U., et al., 1994, A\&A, 284, L9

  Munari, U., et al., 2012,  BaltA 21, 13

  Munari, U., et al., 2014, MNRAS 440, 3402

  Raj, A. et al., 2012, MNRAS 425, 2576

  Shafter, A.~W., et al., 2011, ApJ 727, 50

  Shore, S. N., Gehrz, R. D., 2004, A\&A 417, 695

  van den Bergh S., Younger P.~F., 1987, A\&AS, 70, 125

  Wagner, R. M., et al., 2012, CBET 3136

  Walter F.~M., et al., 2012, PASP, 124, 1057 

  Williams, S.~C., et al., 2013, ApJL, 777, L32

\endreferences 
\clearpage

\end{document}